\documentclass[twocolumn]{aastex63}
\usepackage{graphicx}
\usepackage{amsmath}

\received{XXX}
\revised{YYY}
\accepted{ZZZ}

\newcommand{\chimepsr}{CHIME/Pulsar}
\newcommand{\presto}{PRESTO}
\newcommand{\psrchive}{PSRCHIVE}
\newcommand{\dspsr}{DSPSR}
\newcommand{\psrdada}{PSRDADA}
\newcommand{\tempo}{TEMPO}
\newcommand{\tempotwo}{TEMPO2}
\newcommand{\us}{$\mu s$}
\newcommand{\dmunits}{$\mathrm{pc\,cm^{-3}}$}
\newcommand{\Jtwo}{PSR\,J0218+4232}
\newcommand{\Jfifteen}{PSR\,J1518+4904}
\newcommand{\Jtwenty}{PSR\,J2023+2853}
\newcommand{\pbdottotal}{($\dot{P}_{\rm b})_{\rm obs}$ = $(\dot{P}_{\rm b})_{\rm GR}$ + $(\dot{P}_{\rm b})_{\rm DR}$ + $(\dot{P}_{\rm b})_z$ + $(\dot{P}_{\rm b})_\mu$}
\newcommand{\ef}{}
\newcommand{\cm}{}

\begin{document}
\title{High-cadence Timing of Binary Pulsars with CHIME}

\author[0000-0001-7509-0117]{Chia Min Tan}
    \affiliation{Department of Physics, McGill University, 3600 rue University, Montr\'eal, QC H3A 2T8, Canada}
    \affiliation{The Trottier Space Institute at McGill, 3550 rue University, Montr\'eal, QC H3A 2A7, Canada}
    \affiliation{International Centre for Radio Astronomy Research, Curtin University, Bentley, WA 6102, Australia}

\author[0000-0001-8384-5049]{Emmanuel Fonseca}
    \affiliation{Department of Physics and Astronomy, West Virginia University, Morgantown, WV 26506-6315, USA}
    \affiliation{Center for Gravitational Waves and Cosmology, Chestnut Ridge Research Building, Morgantown, WV 26505, USA}

\author[0000-0002-1529-5169]{Kathryn Crowter}
    \affiliation{Department of Physics and Astronomy, University of British Columbia, 6224 Agricultural Road, Vancouver, BC V6T 1Z1, Canada}

\author[0000-0003-4098-5222]{Fengqiu Adam Dong}
    \affiliation{Department of Physics and Astronomy, University of British Columbia, 6224 Agricultural Road, Vancouver, BC V6T 1Z1, Canada}

\author[0000-0001-9345-0307]{Victoria M. Kaspi}
    \affiliation{Department of Physics, McGill University, 3600 rue University, Montr\'eal, QC H3A 2T8, Canada}
    \affiliation{The Trottier Space Institute at McGill, 3550 rue University, Montr\'eal, QC H3A 2A7, Canada}

\author[0000-0002-4279-6946]{Kiyoshi W. Masui}
    \affiliation{MIT Kavli Institute for Astrophysics and Space Research, Massachusetts Institute of Technology, 77 Massachusetts Ave, Cambridge, MA 02139, USA}
    \affiliation{Department of Physics, Massachusetts Institute of Technology, 77 Massachusetts Ave, Cambridge, MA 02139, USA}

\author[0000-0002-2885-8485]{James W. McKee}
    \affiliation{E. A. Milne Centre for Astrophysics, University of Hull, Cottingham Road, Kingston-upon-Hull, HU6 7RX, UK}
    \affiliation{Centre of Excellence for Data Science, Artificial Intelligence and Modelling (DAIM), University of Hull, Cottingham Road, Kingston-upon-Hull, HU6 7RX, UK}

\author[0000-0001-8845-1225]{Bradley W. Meyers}
    \affiliation{International Centre for Radio Astronomy Research, Curtin University, Bentley, WA 6102, Australia}

\author[0000-0001-5799-9714]{Scott M. Ransom}
    \affiliation{National Radio Astronomy Observatory, 520 Edgemont Rd, Charlottesville, VA 22903, USA}

\author[0000-0001-9784-8670]{Ingrid H. Stairs}
    \affiliation{Department of Physics and Astronomy, University of British Columbia, 6224 Agricultural Road, Vancouver, BC V6T 1Z1, Canada}

\shorttitle{Observing Binary Pulsars with CHIME}
\shortauthors{C. M. Tan et al.}

\begin{abstract}
We performed near-daily observations on the binary pulsars \Jtwo{}, \Jfifteen{} and \Jtwenty{} with the Canadian Hydrogen Intensity Mapping Experiment (CHIME). For the first time, we detected the Shapiro time delay in all three pulsar-binary systems, using only 2--4 years of \chimepsr{} timing data. We measured the pulsar masses to be $1.49^{+0.23}_{-0.20}$ M$_\odot$, $1.470^{+0.030}_{-0.034}$ M$_\odot$ and \cm{$1.50^{+0.49}_{-0.38}$ M$_\odot$} respectively. The companion mass to \Jtwo{} \ef{was} found to be $0.179^{+0.018}_{-0.016}$ \ef{M$_\odot$}. We constrained the mass of the neutron-star companion of \Jfifteen{} to be $1.248^{+0.035}_{-0.029}$ M$_\odot$, using the observed apsidal motion as a constraint on mass estimation. The binary companion to \Jtwenty{} was found to have a mass of \cm{$0.93^{+0.17}_{-0.14}$ M$_\odot$}; in the context of the near-circular orbit, this mass estimate suggests that the companion to \Jtwenty{} is likely a high-mass white dwarf. By comparing the timing model obtained for \Jtwo{} with previous observations, we found a significant change in the observed orbital period of the system of $\dot{P_{\rm b}} = 0.14(2) \times 10^{-12}$\ef{; we determined that this variation arises from ``Shklovskii acceleration" due to relative motion of the binary system, and used this measurement to estimate a distance of $d=(6.7 \pm 1.0)$ kpc to \Jtwo{}.} This work demonstrates the capability of high-cadence observations, enabled by the CHIME/Pulsar system, to detect and refine general-relativistic effects of binary pulsars over short observing timescales.
\end{abstract}

\keywords{instrumentation: interferometers -- methods: observational -- radio continuum: general -- pulsars: general -- techniques: interferometric -- telescopes}

\section{Introduction}
Timing studies of radio pulsars typically depend on the overall time span of observations to measure parameters that quantify apparent or intrinsic spin variations. Until recently, the {\it cadence} of pulsar-timing observations was limited to $\sim$monthly observing rates due to competitive telescope resources. The recent rise of high-cadence pulsar observing programs, such as the UTMOST \citep{jbv+19,lbs+20} and \chimepsr{} \citep{abb+21} experiments, are producing uniquely dense data sets for a majority of the known pulsar population. 

The \chimepsr{} system observes up to 10 sources simultaneously at any given time, allowing for near-daily observations of a selected set of pulsars. This circumstance has resulted in several studies that leverage high-cadence observations to make unique measurements on a growing number of sources. For example, high-cadence observations of the binary PSR J2108+4516 reveal extreme, intra-day variations in the electromagnetic properties of the broadband pulsar signal due to a turbulent wind and/or circumstellar disk of its high-mass companion star \citep{afm+23}. Moreover, high-cadence \chimepsr{} data recently contributed to the refined timing analysis of the binary PSR J0740+6620 for improved mass constraints~\citep{fcp+21}, which directly led to the estimation of its radius using X-ray data obtained by the NICER telescope~\citep{rwr+21, mld+21}. High-cadence \chimepsr{} observations are supporting the analysis of pulsars discovered by the PALFA~\citep{psf+22} and GBNCC~\citep{slr+14} surveys, as well as contributing to the data set developed by NANOGrav for improved detection of the nHz-frequency background of gravitational radiation \citep[e.g.,][]{lam18}. In all of these aforementioned works, the high-cadence nature of \chimepsr{} data affords new views into timing-based phenomena over observing timescales that cannot be achieved with other Northern-hemisphere instruments.

In this study, we perform high-precision timing analyses of three binary pulsars and explore the impact of high-cadence observations obtained with the \chimepsr{} backend. Two of these pulsars -- PSRs J0218+4232 and J1518+4904 -- have long been known and well-studied using archival data that span decades, and thus allow for independent checks of model accuracy. The remaining source, \Jtwenty, is a binary pulsar recently discovered in a survey of the Galactic plane using the Five-hundred Meter Aperture Synthesis Telescope \citep[FAST;][]{hww+21}; to our knowledge, our study below presents the first coherent timing solution of \Jtwenty{} and demonstrates unambiguous period variations due to orbital motion. In Section \ref{sec:obs}, we outline the logistics of our observing program and processing of data. In Section \ref{sec:method}, we outline the procedures used to construct timing models for all three binary pulsars. In Section \ref{sec:results}, we present the results obtained from high-precision timing analyses.  

\section{Observations \& Reduction}
\label{sec:obs}
All data presented in this work were acquired with the CHIME telescope, a static interferometer that digitizes the sky as it drifts within its field of view \citep{abb+22}. The CHIME telescope digitizes raw voltages from 1,024 dual-polarization antennae, sensitive to the 400--800 MHz range, that are mounted on four half-cylinder reflectors that span a $80 \times 100 \  m^2$ area; these digitized samples are then beamformed within the ``FX" correlator to yield up to 10 baseband streams. Beamformed baseband data are phased to user-specified celestial positions in order to track the known radio sources.

These baseband streams are transmitted to the \chimepsr{} backend, a ten-node, real-time computing cluster that enables CHIME to make measurements of radio pulsars and other radio-transient phenomena under user-specified acquisition modes \citep{abb+21}. Each \chimepsr{} node receives one baseband stream, complex-sampled at a rate of 2.56~$\mu$s and for 1,024 frequency channels that span the 400--800 MHz range.

Timing data for our study were obtained using the \chimepsr{} backend in its ``fold'' mode. For all fold-mode observations, input voltages are coherently dedispersed~\citep{hr75}, and then folded using an existing timing model to average individual pulses into 10-second ``subintegrations". Coherent dedispersion is achieved by de-convolving the ISM impulse response function in the frequency domain from the input baseband prior to detection and folding of the radio signal, which removes intra-channel smearing due to pulse dispersion \citep[e.g.,][]{hr75}. We used the \dspsr{} software suite \citep{vb11} for real-time coherent dedispersion and folding with a graphics processing unit mounted onto each \chimepsr{} node.

\cm{We initially took 10 observations of \Jtwenty{} in filterbank mode with the \chimepsr{} backend, as no timing model that incorporates the binary orbital parameter existed at the time we began conducting our measurements. We folded the filterbank data to the spin period and dispersion measure presented in~\cite{hww+21} and found that the apparent spin period of the pulsar changes between observations, indicating a potential binary companion. We modelled the variation in the apparent spin period assuming that they arose from Doppler variations due to unmodeled orbital motion, and obtained an initial binary solution that describes a near-circular orbit with binary period of 0.7 days and projected semi-major axis of 4 lt-s. We then continued to observe \Jtwenty{} in ``fold'' mode using a timing model that includes these orbital parameters.}

The resultant fold-mode data yield time-averaged profiles for all four components of the Stokes polarization vector measured over 10-s subintegrations, a user-defined number of profile bins ($n_{\rm bin}$), and 1,024 frequency channels. We set the profile resolution to be $n_{\rm bin} = 512$ for \Jtwo{}, $n_{\rm bin} = 1,024$ for \Jfifteen{} and $n_{\rm bin} = 256$ for \Jtwenty{}. In the case of \Jtwo{}, its 2.3~ms spin period and the fixed 2.56-$\mu$s baseband sample rate jointly lead an upper limit in allowed profile resolution, such that the largest binary value that can be used to evaluate Stokes profiles for fold-mode observations is 512. The profile resolution of \Jtwenty{} is set at $n_{\rm bin} = 256$, which is the default number set for any CHIME/Pulsar observation. By the time we decided to include \Jtwenty{} in this work, there were approximately 1 year of observations. Hence we decided to not change $n_{\rm bin}$ to keep the dataset consistent.

\subsection{Offline Excision and Reduction} \label{sec:oer}
During post-processing, we used the \psrchive{} package \citep{vdo12} and related utilities for offline cleaning and downsampling of raw timing data. We removed frequency channels that contain persistent radio-frequency interference (RFI) using the fold mode data cleaning tool \textsc{clfd} \citep{mbc+19}, combined with a mask of frequency channels that produce unusable data for the particular day. These masks arise due to variable unavailability of a small fraction of FX-correlator nodes that each produce channelized timeseries for four frequency channels.

We further reduced the RFI-excised timing data by downsampling all spectra from 1024 frequency sub-bands to a smaller number, depending on the S/N of the pulsar, and fully integrating over time. We downsampled the number of frequency sub-bands to $n_{\rm sub} = 32$ for \Jtwo{}, $n_{\rm sub} = 8$ for \Jfifteen{}, and $n_{\rm sub} = 1$ for \Jtwenty{}. The frequency-resolved spectra of \Jtwo{} and \Jfifteen{} provide opportunities for the modeling of dispersion variations, which we enact and discuss below.

\subsection{Generation and Pruning of TOAs} \label{sec:raw}
We used the common technique of cross-correlating Stokes-$I$ profiles with a noiseless ``template" profile in the Fourier domain for determining pulse times of arrivals (TOAs) with high precision \citep{tay92}. Standard templates were generated by aligning and averaging at least 100 days of data using an adequate a timing solution to form a single, representative profile for each pulsar considered in this work. These profiles were then fully averaged over all subintegrations and frequency channels, and de-noised using a wavelet-smoothing algorithm available in \psrchive{}, \textsc{psrsmooth} to produce the standard templates.

As a final step in data preparation, we cleaned all TOAs data sets using two common methods. First, we applied a signal-to-noise (S/N) threshold to each TOA, such that a TOA derived from a Stokes-I profile with S/N $<$ 4 was discarded from the timing analysis presented below. Any remaining TOAs that significantly deviated from timing-model predictions -- due to RFI and/or instrumental corruption whose statistics bypassed the S/N thresholds and RFI-cleaining algorithms -- were manually removed. These TOA-cleaning steps removed 6.8\%, 3.4\%, and 10.7\% of TOAs from the data sets for \Jtwo{}, \Jfifteen{}, and \Jtwenty{}, respectively.

\section{Timing Methods}
\label{sec:method}

An updated timing model based on CHIME/Pulsar data was produced for each pulsar through the weighted least-squared fitting of the TOAs produced (See Section~\ref{sec:raw}) using the TEMPO\footnote{\url{https://tempo.sourceforge.net}} software package. For all timing models, we explored the significance of various astrophysical quantities: astrometric positions and their proper motion; spin frequency and its first derivative; the Keplerian binary parameters and their first-order variations; and the Shapiro delay parameters. The timing models for \Jfifteen{} and \Jtwo{} include a temporal fit of the variation in the electromagnetic dispersion measure (DM) across the timespan. This method is not used for \Jtwenty{} as we only obtained frequency-averaged TOAs that do not allow for DM modelling over time, primarily due to the faintness of the detection on individual observation epochs.

\ef{We also estimated {\it ad hoc} adjustment factors for our raw TOA uncertainties that produced reduced goodness-of-fit ($\chi^2$) statistics of $\sim$ 1. These factors, commonly referred to as ``EFAC" in pulsar-timing literature, characterize the amount of increase in uncertainties needed for TOA residuals to exhibit Gaussian noise largely due to imperfect estimation of the profile template shape \citep[e.g.,][]{aaa+23c}. We found that EFAC values for our data sets varied from 1.2--1.25, indicating that the TOA-estimation algorithm described in Section \ref{sec:obs} largely determined the arrival times and their uncertainties in a robust manner. We did not further explore other separable sources of TOA noise (e.g., ``ECORR") for our sources as such computational efforts are complicated by the large volume of the \chimepsr{} data sets. As shown in \cite{fcp+21}, the use of ``wideband" TOA algorithms \citep[e.g.,][]{pdr14} can reduce the TOA volume by over an order of magnitude and will therefore be crucial for an efficient analysis of noise properties, though we reserve such an effort for future work.} 

\subsection{Models for Orbital Motion}
We used the ELL1 model \citep{lcw+01} to describe the near-circular orbital motion of \Jtwo{} and \Jtwenty{}. For \Jfifteen{}, we used the DD and DDGR models \citep{dd85,dd86} to parameterize periodic and secular orbital variations due to strong-field gravitation in terms of the mass of the companion star ($m_{\rm c}$) and the total system mass ($m_{\rm tot} = m_{\rm p} + m_{\rm c}$, where $m_{\rm p}$ is the mass of the pulsar). 

Both the DDGR and ELL1 models characterize the Keplerian orbits in terms of the orbital period ($P_{\rm b}$) and radial projection of the semi-major axis along the line of sight ($x = a\sin i/c$, where $i$ is the orbital inclination angle), though differ in their treatment of the orbital shape; the DDGR model uses orbital eccentricity ($e$), the argument of periastron ($\omega$), and epoch of periastron passage ($T_0$) to describe ellipticity, while the ELL1 model instead uses the ``Laplace-Lagrange" parameters ($\epsilon_1$, $\epsilon_2$) and the epoch of passage through the ascending-node longitude ($T_{\rm asc}$) to describe O($e$) deviations from circular orbits. The ELL1 eccentricity parameters are related in the following manner:

\begin{align}
    e &= \sqrt{\epsilon_1^2 + \epsilon_2^2}, \\
    \omega &= \tan^{-1}\bigg(\frac{\epsilon_1}{\epsilon_2}\bigg), \\
    T_0 &= T_{\rm asc} + \bigg(\frac{\omega}{2\pi}\bigg)P_{\rm b}.
\end{align}

\noindent In practice, the ELL1 model is most applicable for binary pulsars with timing precision and orbital parameters that mutually satisfy the criterion $\sigma_{\rm TOA, rms} > xe^2$, where $xe^2$ is the amplitude of the O($e^2$) correction to the ELL1 model.

Some binary pulsars eventually exhibit deviations from Keplerian motion due to various particularly extreme properties. Given the short timescale of our observations, the most likely source of ``post-Keplerian" (PK) variations is strong-field gravitation, which gives rise to: apsidal motion ($\dot{\omega}$); orbital decay ($\dot{P}_{\rm b}$); time dilation and gravitational redshift ($\gamma$); and the ``range" ($r$) and ``shape" ($s$) of the Shapiro delay. According to general relativity, these PK effects are functions of the mass and geometric information of the binary system \citep[e.g.,][]{dt92}:

\begin{align}
    \label{eq:omdot}
    \dot{\omega} ={} & 3\bigg(\frac{P_{\rm b}}{2\pi}\bigg)^{-5/3}\frac{(T_\odot m_{\rm tot})^{2/3}}{1-e^2}, \\
    \label{eq:pbdot}
    \dot{P}_{\rm b} ={} & -\frac{192\pi}{5}\bigg(\frac{P_{\rm b}}{2\pi}\bigg)^{-5/3}\frac{T_\odot^{5/3}m_{\rm p}m_{\rm c}}{m_{\rm tot}^{1/3}(1-e^2)^{7/2}} \notag\\
    & \times \bigg(1 + \frac{73}{24}e^2 + \frac{37}{96}e^4\bigg), \\
    \label{eq:gamma}
    \gamma ={} & e\bigg(\frac{P_{\rm b}}{2\pi}\bigg)^{1/3}T_\odot^{2/3}\frac{m_{\rm c}(m_{\rm tot} + m_{\rm c})}{m_{\rm tot}^{4/3}}, \\
    \label{eq:r}
    r ={} & T_\odot m_{\rm c}, \\
    \label{eq:s}
    s \equiv{} & \sin i = x\bigg(\frac{P_{\rm b}}{2\pi}\bigg)^{-2/3}\frac{m_{\rm tot}^{2/3}}{T_\odot^{1/3}m_{\rm c}},
\end{align}

\noindent \ef{where $T_\odot = GM_\odot/c^3 = 4.925490947$ $\mu$s converts the masses to possess units of solar mass}. The DD model directly measures the left-hand side of Equations \ref{eq:omdot}--\ref{eq:s} in a theory-independent manner, while the DDGR model assumes Equations \ref{eq:omdot}--\ref{eq:s} to directly measure $m_{\rm c}$ and $m_{\rm tot}$.

\subsection{Methods for Modeling ISM Variations}
We modeled the DM variation of each pulsar \ef{using} the DMX routine of TEMPO, which estimates a ``local" value of DM within contiguous, non-overlapping time bins that span each data set. The measured DMs over different epochs are then subtracted from the residuals to model the other parameters. For \Jtwo, we set the width of each DMX time bin to be no larger than 5 days; for \Jfifteen, we set the width of DMX time bin to be no larger than 7 days\ef{; and for \Jtwenty{}, we set the DMX width to be no larger than 10 days. These widths were selected based on the variations in the observed trends, but were otherwise chosen arbitrarily.}

\ef{We did not explicitly model or estimate the presence of pulse scatter-broadening in any of the three pulsars subject to our study. This choice was made for two reasons: the use of TEMPO and channelized TOAs does not allow for simultaneous estimation of DM and scattering-timescale parameters; and there is a lack of a scatter-broadening signature in their \chimepsr{} data sets that is distinct from intrinsic profile evolution across the CHIME band. We nonetheless implicitly modeled the composite effects of scattering-broadening and intrinsic profile evolution by fitting for ``frequency-dependent" (FD) variations in our channelized TOAs. These variations were modeled using the FD time delay $\Delta t_{\rm FD} = \sum_i c_i\ln^i\nu$, where $c_i$ is the FD coefficient and $\nu$ is the electromagnetic frequency \citep{nano15}.} 

\ef{When simultaneously estimated with all other timing parameters with TEMPO, we found that at least one FD coefficient was sufficient for \Jtwo{} and the other two pulsars yielded insignificant FD parameters. Only two FD coefficients appear to be statistically significant for \Jtwo{}, but the use of two coefficients leads to clear degeneracy with the DMX parameters. This degeneracy likely arises due to several eras of intra-day variations in DM not being robustly modeling using our 5-day DMX bin widths. We therefore used only one FD coefficient in our final timing model for \Jtwo{}, and manually estimated an EFAC using this ``DMX+FD" timing model in order to correct our TOA uncertainties. The resultant EFAC value estimated from this model likely includes systematic biases from DM mis-estimation and any measurable variations in scatter-broadening, both of which cannot be modeled by the $c_i$ parameters as they are presumed to be constant across each data set. However, the EFAC for \Jtwo{} is only $\sim$ 25\% larger than unity, indicating that the ISM variations are largely captured by the DMX+FD timing model.}

\ef{DMX measurements are useful for evaluating ISM fluctuations across a wide range of length scales, and are often required for any analysis of sub-$\mu$s timing effects like the Shapiro delay and nHz-frequency gravitational radiation \citep{jml+17,sla+23}. The DM variations observed in \Jtwo{} are particularly well-resolved as shown in Figure \ref{fig:J0218_DMX}, such that a lack of DMX evaluation would contribute substantial ``red" noise to our timing measurements and dominate the best-fit uncertainties. We reserve an analysis of the DMX timeseries for \Jtwo{} and other sources for a forthcoming census of low-frequency DM measurements with \chimepsr{} \citep{mck24}.}

\subsection{Goodness-of-Fit Grids of Shapiro Delay Parameters} \label{sec:goodnessoffit}

With the timing model in hand, we used the Bayesian $\chi^2$-gridding method developed by \cite{sna+02} to better determine the statistics of the Shapiro delay parameters for each pulsar. In this method, we iterated over the Shapiro delay parameters in a $100 \times 100$ grid of ($m_{\rm c}$, $\cos i$) values. At each ($m_{\rm c}$, $\cos i$) grid point, the timing model of each pulsar is refitted while fixing the values of the Shapiro delay parameters. A 2-D probability density function (PDF), $p(m_{\rm c}, \cos i|{\rm data})$, is then generated by comparing the reduced $\chi^2$ values obtained at different Shapiro delay parameter value pairs.

The two-dimensional PDFs obtained for each pulsar were then used to constrain the Shapiro-delay parameters and the mass of the pulsar\ef{. For the estimates of $m_{\rm p}$}, we used the Keplerian mass function ($m_{\rm f}$), where

\begin{align}
\label{eq:mass_function}
m_{\rm f} &= \frac{\ef{(}m_{\rm c} \sin i\ef{)}^3}{m_{\rm tot}^2} \ef{= \frac{4\pi^2}{T_\odot}\frac{x^3}{P_{\rm b}^2}},
\end{align}

\noindent \ef{in order} to enact the transformation in PDF variables:

\begin{align}
    p(m_{\rm c}|{\rm data}) &= \int_0^\infty p(m_{\rm c}, \cos i|{\rm data})d(\cos i), \\
    p(m_{\rm p}|{\rm data}) &=  \notag\\
    \int_0^\infty \int_0^\infty & p(m_{\rm c}, \cos i|{\rm data})\delta(\Delta m_{\rm p}) d(\cos i)d(m_{\rm c}), \\
    p(\cos i|{\rm data}) &= \int_0^\infty p(m_{\rm c}, \cos i|{\rm data})d(m_{\rm c}),
\end{align}

\noindent where $\delta(x)$ is the Dirac delta function, and $\Delta m_{\rm p} = m_{\rm p} - (\sqrt{(m_{\rm c}\sin i)^3/m_{\rm f}} - m_{\rm c})$. We computed the cumulative distribution functions (CDFs) for each of the three above parameters to derive the median, 68.3\% and 95.4\% credible-interval values for robust estimates of the relativistic parameters and their statistical uncertainties.

\subsection{Constrained Shapiro-delay Grid for \Jfifteen}
\Jfifteen{} exhibits apsidal motion that has long been known to be consistent with the predictions of general relativity \citep{nst96,jsk+08}, as given in Equation \ref{eq:omdot}. Since $\dot{\omega}$ is a function of $m_{\rm tot}$, then its statistical significance can be used as a weight on grid-based estimates of ($m_{\rm c}$, $\cos i$) and further delimit regions of preferred values, as has been done for other high-eccentricity systems. We therefore performed an additional $\chi^2$-grid calculation for \chimepsr{} timing data of \Jfifteen where, at each grid point, we held values of $m_{\rm c}$, $m_{\rm tot}$, \ef{and $\dot{\omega}$} fixed, using \ef{Equation \ref{eq:mass_function}} to compute and fix $\sin i$.

We chose to perform grid calculations for \Jfifteen{} in the ($m_{\rm c}$, $m_{\rm tot}$) phase space due to the high statistical significance of $\dot{\omega}$. After extensive analysis, we found that a grid over the ($m_{\rm c}$, $m_{\rm tot}$) phase space produces better constraints on the probability densities compared with that from the traditional ($m_{\rm c}$, $\cos i$) space. These calculations ultimately produced a map of probability density $p(m_{\rm c}, m_{\rm tot}|{\rm data})$. Using this map, we derived estimates of the Shapiro-delay parameters by marginalizing over the appropriate coordinate axes:

\begin{align}
    p(m_{\rm c}|{\rm data}) &= \int_0^\infty p(m_{\rm c}, m_{\rm tot}|{\rm data})\delta(\Delta\dot{\omega})d(m_{\rm tot}), \\
    p(m_{\rm p}|{\rm data}) &= \int_0^\infty \int_0^\infty p(m_{\rm c}, m_{\rm tot}|{\rm data}) \nonumber \\ &\phn{} \times \delta(\Delta m_{\rm p}) d(m_{\rm tot})d(m_{\rm c}), \\
    p(\cos i|{\rm data}) &= \int_0^\infty \int_0^\infty p(m_{\rm c}, m_{\rm tot}|{\rm data}) \nonumber \\ &\phn{} \times \delta(\Delta \cos i) d(m_{\rm tot})d(m_{\rm c}),
\end{align}

\noindent where $\Delta \cos i = \cos i - \sqrt{1 - (m_{\rm f}m_{\rm tot}^2 / m_{\rm c}^3)^{2/3}}$.

\begin{figure}
    \centering
    \includegraphics[width=\linewidth]{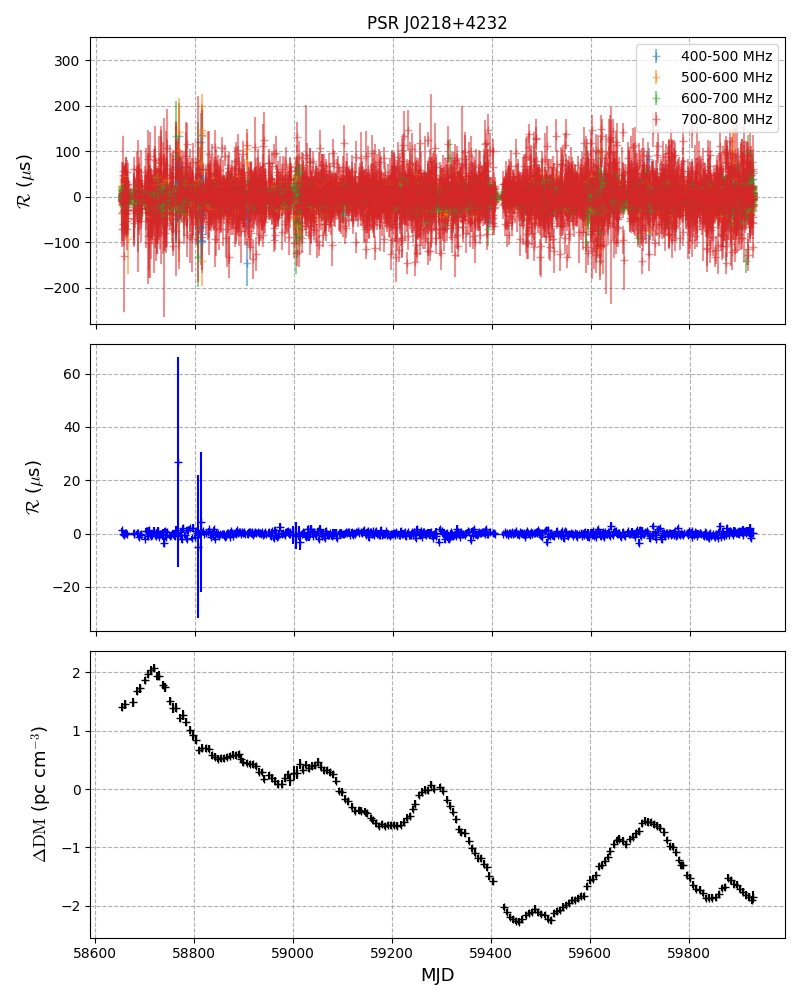}
    \includegraphics[width=\linewidth]{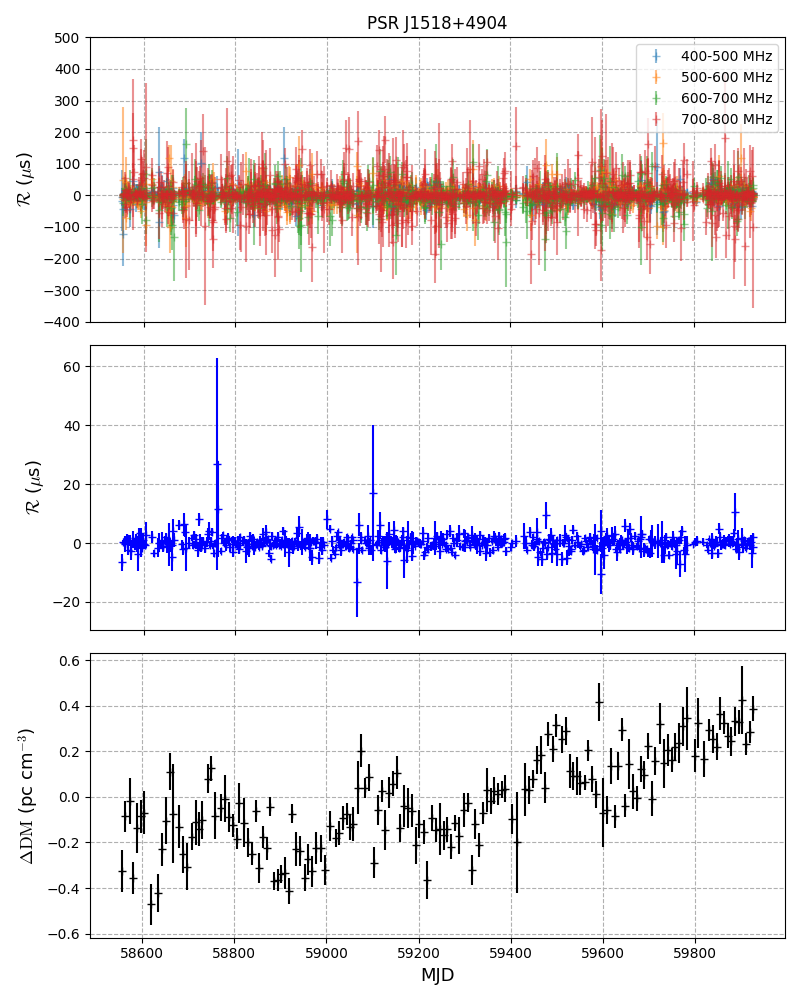}
    \caption{\cm{Frequency-resolved and frequency-averaged-timing residuals (top and middle panels, respectively),} and dispersion measure variation \cm{(bottom panels)} of \Jtwo{} and \Jfifteen{} over the observing span, as measured using the DMX method. While there is evidence for an annual variation in the DM of \Jfifteen{}, the ecliptic latitude of \Jfifteen{}, $\beta = 63^\circ$, is too large for electron-density variations from the Sun to appear prominently in our DM timeseries.}
    \label{fig:J0218_DMX}
\end{figure}

\begin{figure}
    \centering
    \includegraphics[width=\linewidth]{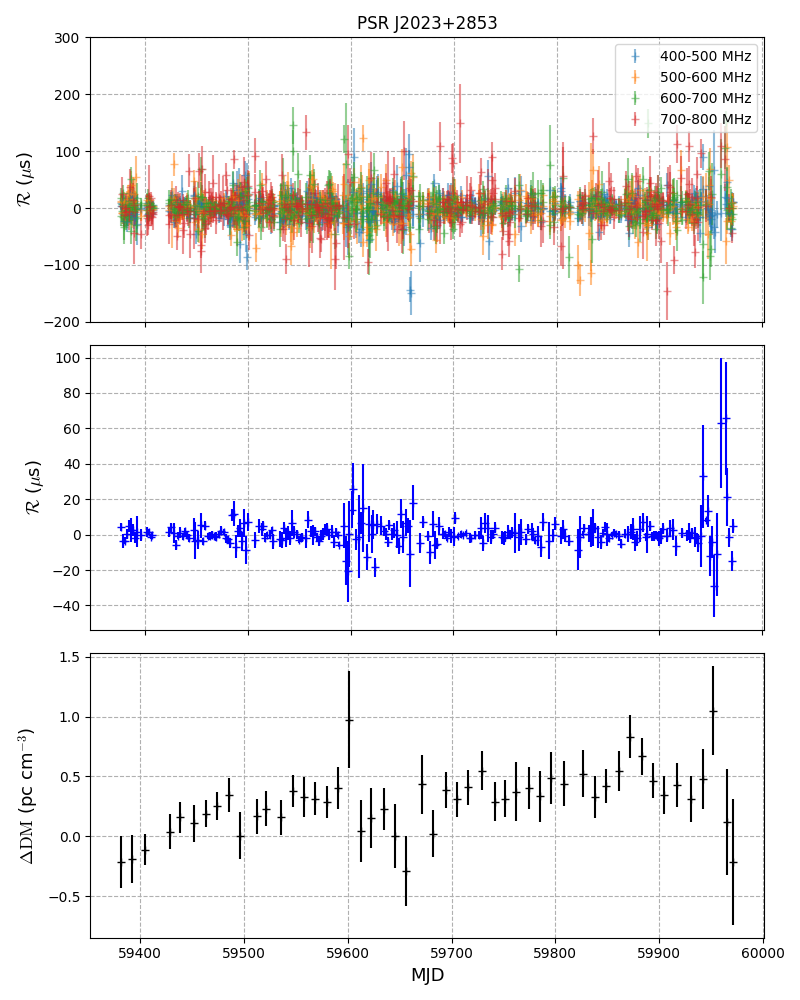}
    \caption{\cm{Frequency-resolved and-frequency averaged timing residuals (top and middle panels, respectively),} and dispersion measure variation \cm{(bottom panel)} of \Jtwenty{} over the observing span, as measured using the DMX method.}
    \label{fig:J2023_DMX}
\end{figure}

\begin{figure}
    \centering
    \includegraphics[width=\linewidth]{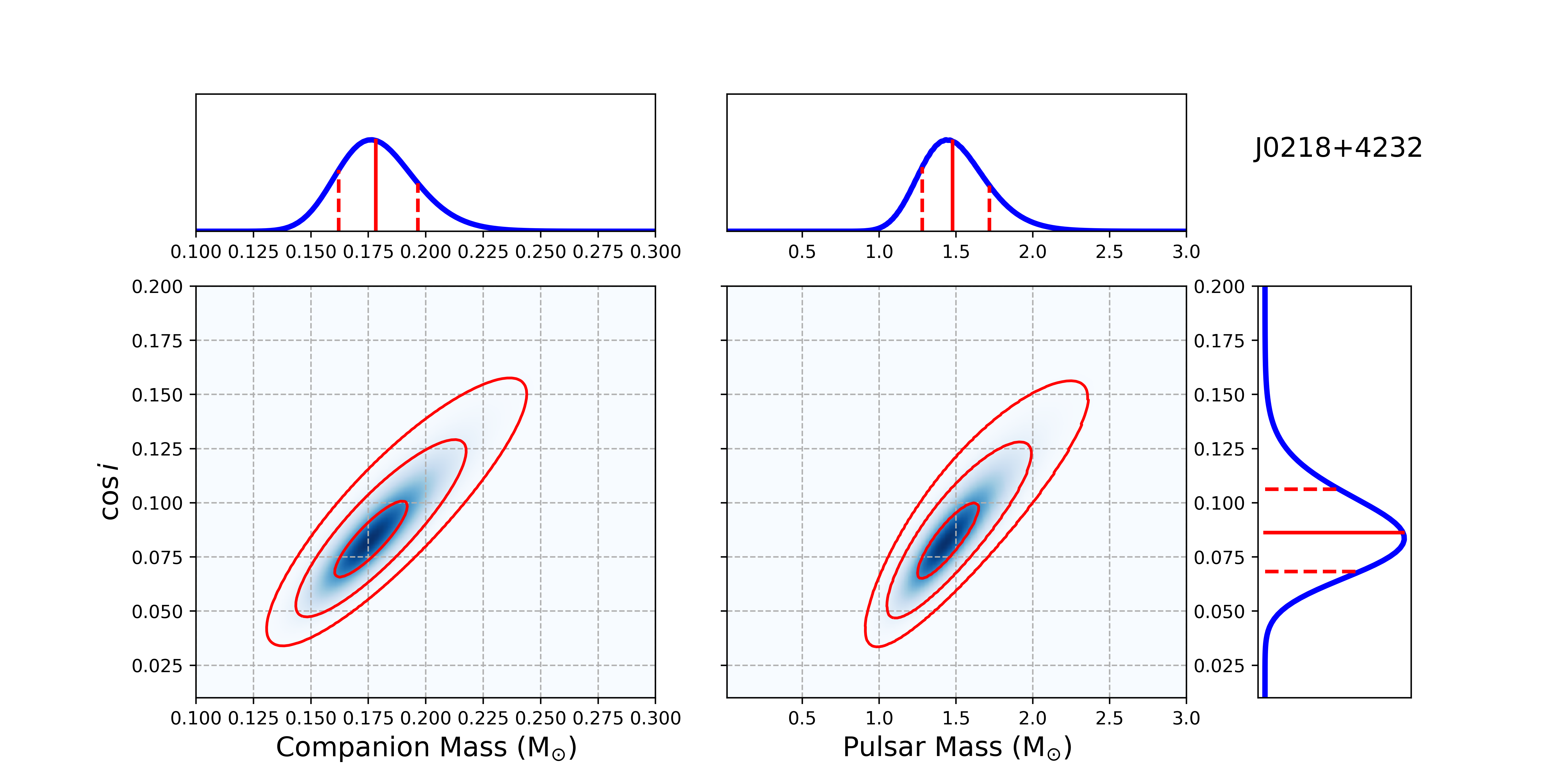}
    \includegraphics[width=\linewidth]{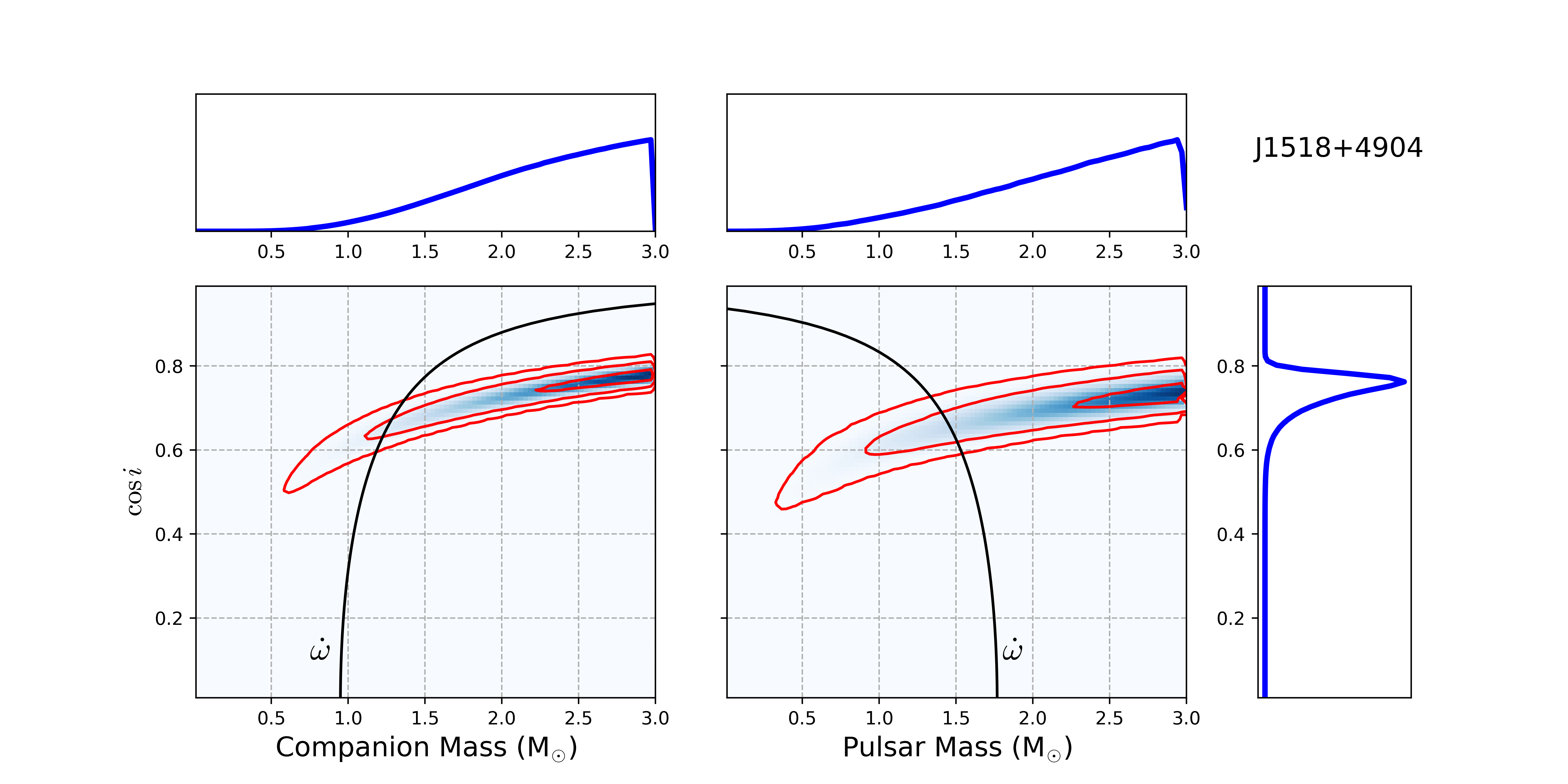}
    \includegraphics[width=\linewidth]{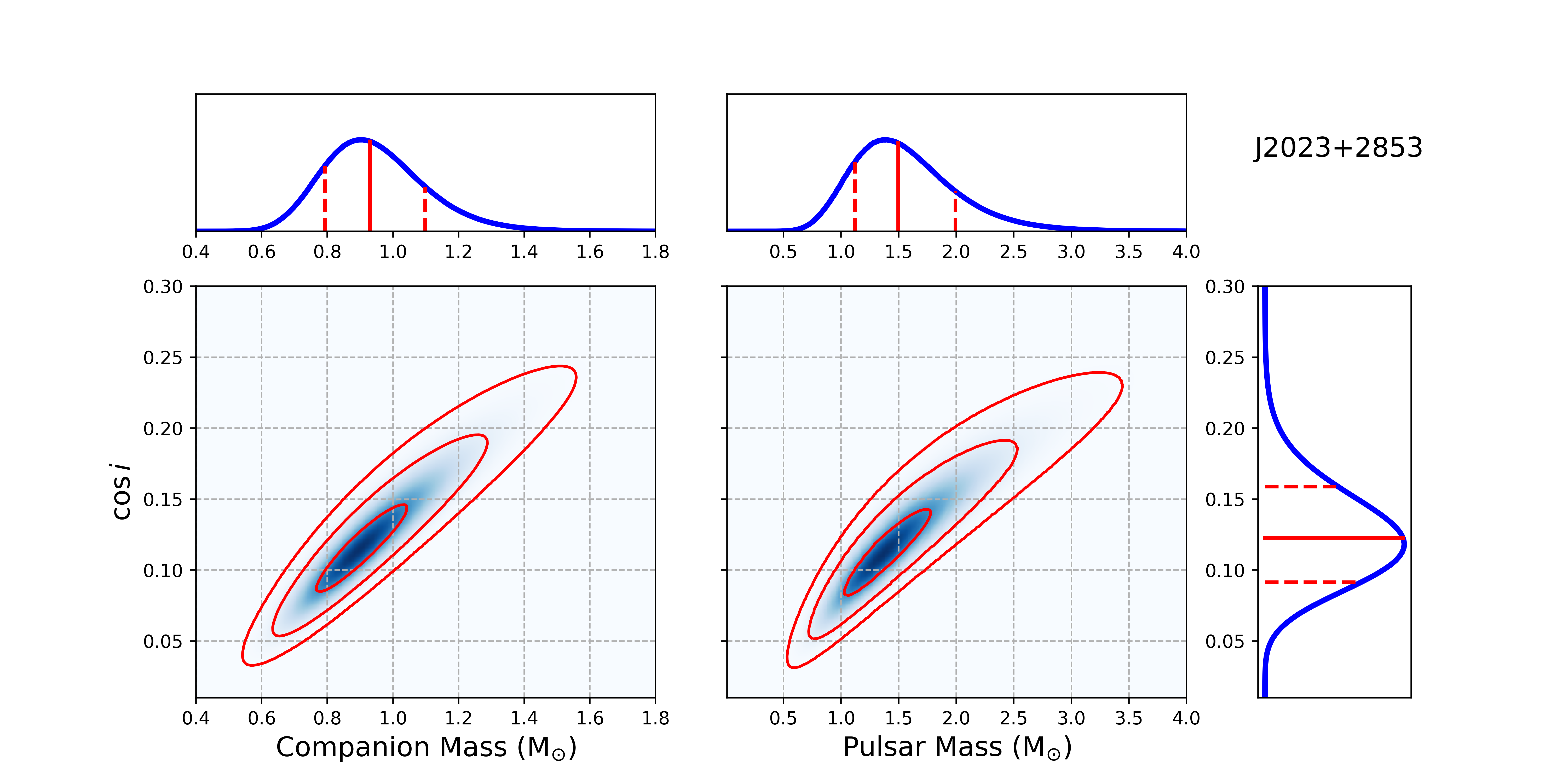}
    \caption{The probability density maps of Shapiro-delay parameters for PSRs J0218$+$4232\ef{, J1518$+$4904,} and J2023$+$2853. The red contours in each map represent regions of probability that contain 68.3\%, 95.4\%, and 99.3\% credibility. In the outer panels, red-solid lines denote median values while the red-dashed lines represent 68.3\% credible intervals. \ef{For \Jfifteen{}, the black curves in each map represent the range of allowed masses inferred by the observed $\dot{\omega}$ when assuming that general relativity describes the apsidal motion; the precision in our measurement of $\dot{\omega}$ is so high that the two curves appear as a single line in each map. No credible intervals are presented in the outer panels for \Jfifteen{} as the majority of  probability density is not fully encapsulated in the map bounds.}}
    \label{fig:SDgrids}
\end{figure}

\begin{figure}
    \centering
    \includegraphics[width=\linewidth]{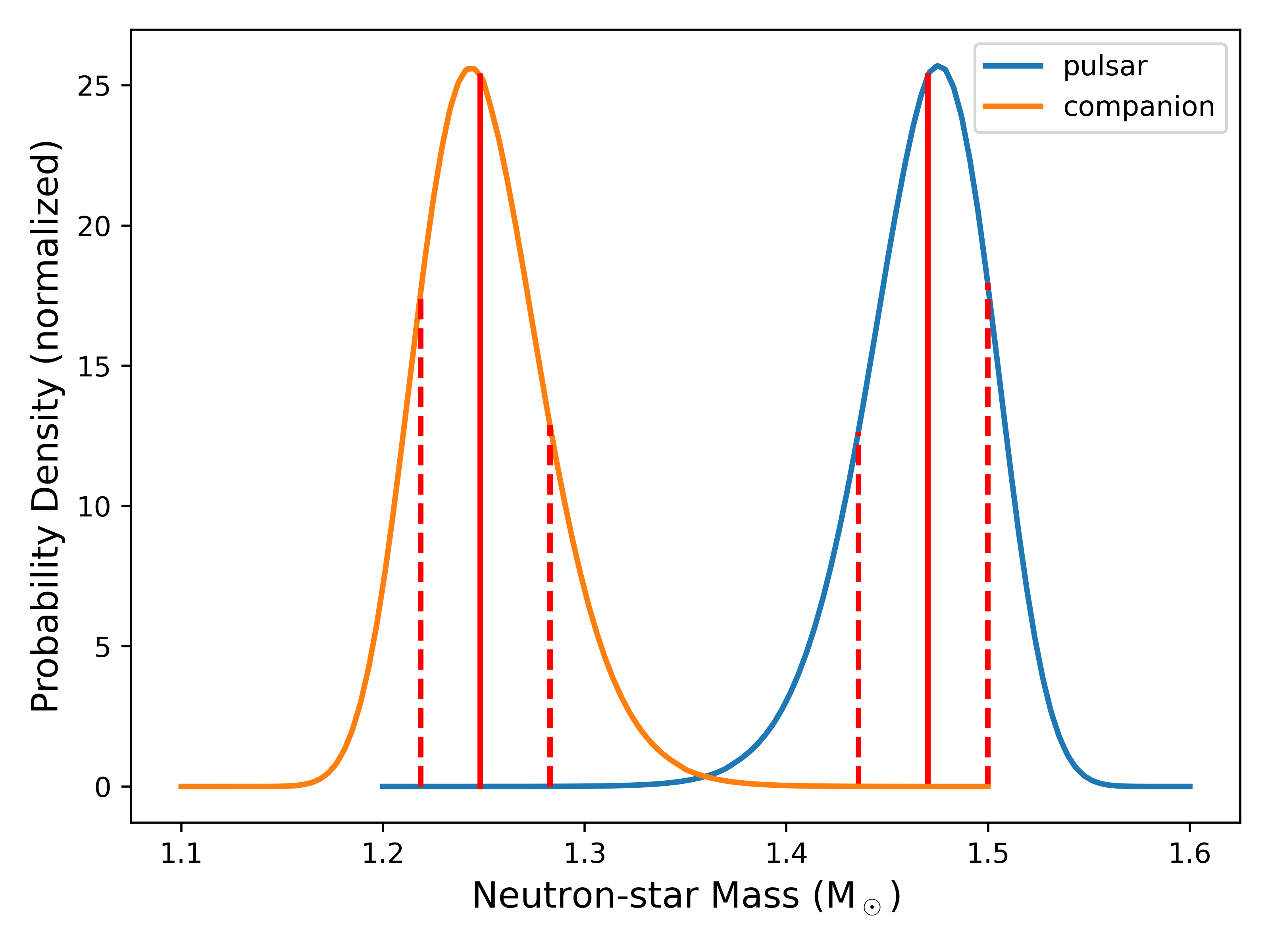}
    \caption{The \ef{constrained} probability density of the neutron-star masses in the \Jfifteen{} system, derived from \ef{posterior} PDFs of the Shapiro-delay parameters \ef{computed while using} the observed apsidal motion as a constraint on the likelihood function. The red solid line is the median value while the red-dashed lines represent 68.3\% credible intervals of the masses.}
    \label{fig:J2023_SDgrid}
\end{figure}



\begin{table*}
\centering
\caption{Best-fit parameters and derived quantities for binary MSPs. \label{tab:timing_solution}}
\begin{tabular}{lccc}
\hline\hline
\multicolumn{4}{c}{Global Parameters} \\
\hline
Pulsar name\dotfill & \Jtwo & \Jfifteen & \Jtwenty\\
Reference epoch (MJD)\dotfill & 58917 & 59134.411 & 59674 \\
Observing timespan (MJD)\dotfill & 58652-59930 & 58552-59929 & 59376-59930 \\
Solar System Planetary Ephemeris \dotfill & DE440 & DE440 & DE440 \\
Binary model \dotfill & ELL1 & DDGR & ELL1 \\
Clock Standard \dotfill & TT(BIPM2021) & TT(BIPM2021) & TT(BIPM2021) \\
Number of Sub-bands\dotfill & 32 & 8 &\cm{8}\\
\cm{Width of DMX time bin (days)\dotfill} & \cm{5} & \cm{7} & \cm{10}\\
TOA Uncertainty Adjustment Factor (EFAC) \dotfill & 1.2 & 1.25 & \cm{1.25}\\
\hline
\multicolumn{4}{c}{Timing Solution \& Best-fit Metrics} \\
\hline
Right ascension (J2000), $\alpha$ (h:m:s) \dotfill & 02:18:06.362483(15) & 15:18:16.797784(12) & \cm{20:23:21:06320(7)} \\
Declination (J2000), $\delta$ (d:m:s) \dotfill & 42:32:17.3417(3) & 49:04:34.08325(12) & \cm{28:53:41.4423(7)}\\
Proper motion in Right ascension, $\mu_{\alpha}\cos\delta$ ($\rm mas\,yr^{-1}$)\dotfill & 5.39(11) & $-0.72(12)$ & \cm{$-2.6(19)$} \\
Proper motion in Declination, $\mu_{\delta}$ ($\rm mas\,yr^{-1}$)\dotfill & 1.4(2) & $-8.57(11)$ & \cm{$-6(4)$}\\
Parallax, $\varpi$ (mas)\dotfill & --- & 4(2) & \cm{---}\\
Pulse frequency, $\nu$ (s$^{-1}$)\dotfill & 430.461056366647(6) &  24.4289797496997(2) & \cm{88.269785050593(8)}\\
First pulse frequency derivative, $\dot{\nu}$ ($10^{-15}$\, s$^{-2}$)\dotfill & $-14.34202(16)$ & $-0.016197(10)$ & \cm{$-0.212(3)$}\\
Dispersion measure, DM (\dmunits{})\dotfill & 61.233749 & 11.611711 & 22.75 \\
Orbital period, $P_{\rm b}$ (days)\dotfill & 2.02884608487(5) & 8.63400496116(15) & \cm{0.71823040745(4)}\\
Projected semi-major axis, $x$ (lt-s)\dotfill & 1.98443196(5) & 20.03942440(11) & \cm{4.0022194(4)}\\
Epoch of periastron passage, $T_0$ (MJD) \dotfill & --- & 59125.99829199(5) & ---\\
Orbital eccentricity, $e$\dotfill & --- & 0.249484383(9) & --- \\
Longitude of periastron, $\omega$ (deg) \dotfill & --- & 342.745426(2) & ---\\
Epoch of ascending node passage, $T_{ASC}$ (MJD) \dotfill & 58915.445027883(14) & --- & \cm{59390.14830429(2)}\\
First Laplace-Lagrange parameter, $\epsilon_{1}$ (10$^{-5}$) \dotfill & 0.39(2) & --- & \cm{1.11(2)}\\
Second Laplace-Lagrange parameter, $\epsilon_{2}$ (10$^{-5}$) \dotfill & 0.50(2) & --- & \cm{0.73(2)}\\
\cm{First order frequency dependent variation in channelised TOAs} & \cm{$-1.54(15) \times 10^{-5}$} & --- & ---\\
Number of TOAs \dotfill & 29889 & 7412 & \cm{2623}\\
$\chi^2$\dotfill & 30192.25 & 7458.33 & \cm{2690.70}\\
Degrees of freedom \dotfill & 29675 & 7241 & \cm{2560}\\
Reduced $\chi^2$\dotfill & 1.02 & 1.03 & \cm{1.05}\\
RMS timing residual (\us)\dotfill & 6.458 & 4.906 & \cm{12.395}\\
\hline
\multicolumn{4}{c}{Gridded Estimates of the Masses and Orbital Inclination} \\
\hline
Mass of the companion, $m_{\rm c}$ (M$_\odot$) \dotfill & $0.179^{+0.018}_{-0.016}$ & $1.248^{+0.035}_{-0.029}$ & \cm{$0.93^{+0.17}_{-0.14}$}\\
Total mass of the system, $m_{\rm tot}$ (M$_\odot$) \dotfill & --- & $2.7186 (7)$ & --- \\
Cosine of the inclination angle, $cos\ i$ \dotfill & $0.086^{+0.020}_{-0.018}$ & --- & \cm{$0.123^{+0.036}_{-0.032}$}\\
\hline
\multicolumn{4}{c}{Derived Quantities} \\
\hline
Mass function, $m_{\rm f}$ (M$_\odot$) \dotfill & \cm{0.0020384176(2)} & \cm{0.115908122(2)} & \cm{0.13343133(4)}\\
Mass of the pulsar, $m_{\rm p}$ (M$_\odot$) \dotfill & $1.49^{+0.23}_{-0.20}$ & $1.470^{+0.030}_{-0.034}$ & \cm{$1.50^{+0.49}_{-0.38}$} \\ 
Inclination angle of the binary system, $i$, (degrees) \dotfill & $85.1^{+1.0}_{-1.2}$ & $49.6^{+1.6}_{-1.8}$ & \cm{$83.0^{+1.8}_{-2.0}$}\\
Rate of advance of periastron, $\dot{\omega}$ (deg yr$^{-1}$) \dotfill & --- & 0.011373(2) & ---\\
Time-dilation and gravitational-redshift parameter, $\gamma$ \dotfill & --- & $0.00463^{+0.00017}_{-0.00014}$ & ---\\
Relativistic orbital decay, $\dot{P_{\rm b}}$ ($10^{-12}$) \dotfill & --- & $-0.001176(5)$ & ---\\
\hline                                                                                                
\end{tabular}%
\end{table*}

\section{Results and Discussion} 
\label{sec:results}






We obtained {\ef{best-fit}} timing models for all three binary pulsars using TEMPO and the methods described in Section \ref{sec:method}. The best-fit timing residuals and available DMX data for all binary pulsars are presented in Figure \ref{fig:J0218_DMX}. A summary of each timing model, derived parameters, and best-fit statistics is presented in Table \ref{tab:timing_solution}, while fixing the Shapiro delay values to the median values obtained from the $\chi^2$ grids described in Section~\ref{sec:goodnessoffit}.

\subsection{New Shapiro-delay Measurements}

For the first time, we detected the Shapiro time delay in all three binary pulsar systems. The near-daily observing cadence has clearly led to desirable coverage of the orbital phase over a 2--3-yr timescale; the lack of detection in prior work on PSRs J0218$+$4232~\citep{ndf+95, dcl+16, pdd+19} and J1518$+$4904~\citep{jsk+08} was likely impacted by low-density orbital phase coverage despite their several-decade timespan.

We consider the Shapiro delay in \Jfifteen{} to be detected since total degeneracy of the ($m_{\rm c}$, $\cos i$) parameters -- regions where probability density is non-zero as $\cos i$ tends to unity -- is statistically disfavored within the range of plausible companion-mass values. However, the Shapiro delay in the \Jfifteen{} system is nonetheless weakly detected as the marginalized constraints on $m_{\rm c}$ and $\cos i$ produce a weak constraint on $m_{\rm p}$. As a standalone effect, the current Shapiro delay in \Jfifteen{} therefore offers no meaningful constraining power to the mass and geometry of the binary system, other than to restrict likely values of $i$. However, the use of $\dot{\omega}$ as a statistical weight in the $\chi^2$-grid calculation leads to substantial improvements in the constraints of physical parameters. We found that $m_{\rm p} = 1.470^{+0.030}_{-0.034}$ M$_\odot$ and $m_{\rm c} = 1.248^{+0.035}_{-0.029}$ M$_\odot$ (68.3\% credibility), under the assumption that general relativity describes the observed apsidal motion. 

Our estimates of $m_{\rm p}$ for PSRs J0218$+$4232 and J2023$+$2853 -- $m_{\rm p} = 1.48^{+0.24}_{-0.20}\textrm{ M}_\odot$ and $m_{\rm p} = 1.42^{+0.40}_{-0.32}\textrm{ M}_\odot$, respectively -- are broadly consistent with values observed for the recycled-pulsar population, though additional observations will further strengthen the constraints for delineating whether these pulsars lie on the high end of the neutron-star mass spectrum. The estimates of the neutron-star masses found in the \Jfifteen{} system are consistent with masses found in other DNS systems \citep[e.g.,][]{spr10}.

\subsection{Discussion on individual sources} 

\subsubsection{\Jtwo{}}

The properties of the \Jtwo{} system, first discovered by \cite{ndf+95}, were most recently modelled by~\cite{dcl+16} and later~\cite{pdd+19} as part of the International Pulsar Timing Array (IPTA) project. No apparent detection of the Shapiro delay was made in either of their extended data sets. By contrast, we were able to successfully detect the Shapiro-delay signature in the \Jtwo{} system due to the near-daily cadence of our \chimepsr{} data set. \ef{The measurement of $\cos i$ is consistent with constraints placed on orbital geometry and spin orientation of \Jtwo{} derived from low-frequency polarimetry \citep{stc99}, and suggests that the misalignment between the axes of orbital angular momentum and pulsar spin are nearly orthogonal.}  

The statistical significance of the Shapiro delay remains high even when incorporating hundreds of degrees of freedom from the DMX model into our timing analysis. However, when included \ef{as a} free parameter, the best-fit measured proper motion in declination was found to be inconsistent with the measurements by both~\cite{dcl+16} and~\cite{pdd+19}, as well as a Very Long Baseline Interferometry (VLBI) measurement by~\citep{dyc+14}, both in magnitude and sign. Moreover, we found that fits of the timing signature due to parallax yielded negative values with apparent statistical significance. We re-modeled the properties of \Jtwo{} after fixing the parallax to zero and proper motion to the model-independent values determined by~\cite{dyc+14} of $\mu_{\alpha} \cos\delta = 5.35$ mas yr$^{-1}$ and $\mu_{\delta} = -3.74$ mas yr$^{-1}$, and found that the best-fit DM variations of the pulsar changed significantly. We suspect that these discrepancies arise due to several reasons: the sensitivity of our data set on \Jtwo{} to intra-day DM variations; our observations being scheduled such that the pulsar is observed on the same local sidereal time every day; and that the binary orbit of \Jtwo{} is slightly more than exactly 2 days, resulting in the variation of the orbital Doppler shift being absorbed into the DM-variation modelling. This apparent DM variation is found to be degenerate with the astrometric parameters and manifests itself as an offset to the measured proper motion in declination. These issues do not arise with the other pulsars analyzed in our study.

We nonetheless found that the difference in the proper motion used to model the binary system resulted in an insignificant difference in the measured Shapiro Delay parameters. We found $m_\textrm{c}$ $=0.182^{+0.019}_{-0.017}$ M$_\odot$ and $\cos\ i = 0.0911^{+0.020}_{-0.019}$ after fixing the proper motion of the system to be the values obtained by~\cite{dyc+14}. These values remain consistent at the 68.3\% credibility level with values derived from a timing model that sets the proper motion of the system as free parameters.

The \chimepsr{} timing model provides a high-cadence snapshot of the \Jtwo{} orbit that may have evolved over the decades since its discovery. We therefore compared our values of the orbital elements with those most recently published for the second data release of IPTA \citep{pdd+19}; the IPTA data set for \Jtwo{} spans $\sim$17.5 years and consists of TOAs collected with a variety of European observatories. As the timing model from~\cite{pdd+19} does not include any Shapiro-delay modelling, we first downloaded the publicly available data from their work and refit the release model for \Jtwo{} after incorporating the Shapiro-delay parameters as degrees of freedom. This refitting was done to ensure robust comparisons of models, and to avoid potential biasing of orbital parameters due to the presence of an unmodeled Shapiro delay \citep{fw10}. 

Using this new IPTA model, we found that $x$ has remained the same between the two orbital models to within $1\sigma$, where $\sigma$ is the larger of the two statistical uncertainties in the fit parameter. We inferred an approximate rate of change in $x$ to be $\dot{x} \approx \Delta x / \Delta T_{\rm asc} = -5(4)\times10^{-15}$ s/s. This limit on $\dot{x}$ is consistent with the maximum value of $\dot{x}$ due to the changing of orientation due to proper motion: given our estimate of $i$ and that $(\dot{x})_{\rm max} = x\mu |\cot{i}|$ \citep{nss01}, we found that $|\dot{x}|_{\rm max} \sim 10^{-16}$ s/s.

By contrast, we found that there is a significant change in $P_{\rm b}$ between our \chimepsr{} timing model and the Shapiro-delay incorporated model derived from the IPTA data set from~\cite{pdd+19}. We computed the change in $P_{\rm b}$ between the two epochs to be $\dot{P_{\rm b}} \approx \Delta P_{\rm b}/\Delta T_{\rm asc} = 0.14(2) \times 10^{-12}$, where the uncertainty is dominated by those estimated for $P_{\rm b}$. 

The fact that $\dot{P_{\rm b}}$ is positive suggests this variation is not dominated by decay due to the emission of gravitational waves from the system. Instead, this rate of change most likely arises from variations in the Doppler shift due the relative motion between \Jtwo{} and Earth. This motion results in contributions of $\dot{P}_{\rm b}$ to the total, observed value, such that \pbdottotal, where $(\dot{P}_{\rm b})_{\rm GR}$ is general-relativistic orbital decay rate presented in Equation \ref{eq:pbdot} and the additional terms are defined as follows \citep{shk70,kg89,dt92,nst96}: 

\begin{align}
    \label{eq:pbdot_DR}
    (\dot{P}_{\rm b})_{\rm DR} &= -P_{\rm b}\cos b \bigg(\frac{\Theta_0^2}{cR_0}\bigg)\bigg(\cos l + \frac{\kappa^2}{\sin^2l + \kappa^2}\bigg), \\
    \label{eq:pbdot_z}
    (\dot{P}_{\rm b})_{\rm z} &= -1.08\times10^{-19}\frac{P_{\rm b}}{c} \notag\\
    & \times \bigg(\frac{1.25z}{[z^2 + 0.0324]^{1/2}} + 0.58z\bigg)\sin b, \\
    \label{eq:pbdot_mu}
    (\dot{P}_{\rm b})_\mu &= \frac{\mu^2d}{c}P_{\rm b}.
\end{align}

\noindent The various terms in Equations \ref{eq:pbdot_DR}--\ref{eq:pbdot_mu} are defined as follows: $b$ and $l$ are the Galactic latitude and longitude, respectively; $\Theta_0$ and $R_0$ are the Galactrocentric circular-speed and distance parameters for the Solar-System barycenter, respectively; $\kappa = (d/R_0)\cos b - \cos l$, where $d$ is the distance to the pulsar; $z = d\sin b$ is the projected vertical distance of the pulsar-binary system from the Galactic plane; and $\mu = \sqrt{\mu_\alpha^2 + \mu_\delta^2}$ is the magnitude of proper motion. We assumed $\Theta_0$ = 236.9(4.2) km s$^{-1}$ and $R_0$ = 8.178(26) kpc as recently determined by \citet{aab+19}.

By modelling $(\dot{P}_{\rm b})_{\rm obs}$ in terms of Equations \ref{eq:pbdot} and \ref{eq:pbdot_DR}--\ref{eq:pbdot_mu}, we found that large values of $d$ are needed to produce our derived value. \ef{We performed a Monte Carlo analysis of $d$ by randomly sampling the \{$m_{\rm p}$, $m_{\rm c}$, $\Theta_0$, $R_0$, $\mu$, $(\dot{P}_{\rm b})_{\rm obs}$\} parameters, based on their best available uncertainties that we assumed to be Gaussian\footnote{\ef{While our Shapiro-delay analysis shows that the posterior PDFs for $m_{\rm p}$ and $m_{\rm c}$ are non-Gaussian, the $(\dot{P}_{\rm b})_{\rm GR}$ term is subdominant to all other sources of variation in $P_{\rm b}$ that therefore has no bearing on our Monte-Carlo estimate of $d$.}}, and computing $d$ using Equations \ref{eq:pbdot} and \ref{eq:pbdot_DR}--\ref{eq:pbdot_mu} for each combination of parameter values. Our estimate of $(\dot{P}_{\rm b})_{\rm obs}$ from this Monte Carlo analysis corresponds to $d = (6.7 \pm 1.0)$ kpc, where the statistics reflect the mean and standard deviation of the resultant distribution in $d$. This distance is} statistically consistent with the VLBI distance determined by \cite{dyc+14}, $d = 6.3$ kpc, where the combined effects give a predicted estimate of $\dot{P_{\rm b}} \approx 0.12 \times 10^{-12}$ for this large distance. Conversely, a distance to the pulsar of $d = 3.15$ kpc, as suggested by~\cite{vl14}, would yield an expected $\dot{P_{\rm b}} \approx 0.06 \times 10^{-12}$. This latter value of $d$ is inconsistent with our estimate of $(\dot{P}_{\rm b})_{\rm obs}$ at the $4\sigma$ level. Otherwise, the spectroscopic analysis of the white dwarf companion~\citep{bkk03} places the binary pulsar system at a distance of $d = 4$ kpc for a white dwarf model that is consistent with the measured mass. This gives an expected $\dot{P_{\rm b}} \approx 0.08 \times 10^{-12}$, which differs by $3\sigma$ from our measured $\dot{P_{\rm b}}$ value.

While the difference between our measured $\dot{P_{\rm b}}$ value and expected $\dot{P_{\rm b}}$ from several distance estimates is tantalising, the specific form of Equation \ref{eq:pbdot_z} assumes a model of vertical acceleration due to the surface-mass distribution of the Galactic disk estimated by \cite{kg89}. We separately utilized an alternative model of the Galactic surface-mass density developed by \cite{hf00,hf04}, which predicts that 

\begin{align}
    \label{eq:pbdot_z2}
    (\dot{P}_{\rm b})_{\rm z} &= -\bigg[2.27|z|_{\rm kpc} + 3.68(1-\exp[-4.61|z|_{\rm kpc}])\bigg] \nonumber \\
    &\phn{}\times 10^{-11}\bigg(\frac{P_{\rm b}|\sin b|}{c}\bigg)
\end{align}

\noindent and has been used in recent pulsar-timing analyses \citep[e.g.,][]{lwj+09,zsd+15}. Both Equations \ref{eq:pbdot_z} and \ref{eq:pbdot_z2} predict that $(\dot{P}_b)_z \approx-10^{-14}$ s/s for a distance range $3.15 < d < 6.3$ kpc, which are subdominant to the ``Shklovskii acceleration" defined in Equation \ref{eq:pbdot_mu}. The preference for a large distance to \Jtwo{} therefore remains, regardless of the choice in a model of $(\dot{P}_b)_z$. However, these models likely lose accuracy for large $z$; further development of viable Galactic-acceleration models \citep[e.g.,][]{ccl+21} will therefore improve the robustness of our timing-based constraint on $d$ for \Jtwo.

\subsubsection{\Jfifteen{}}

The most recent analysis of the \Jfifteen{} binary system was conducted by \cite{jsk+08}, using a data set that spanned over 12 years. The properties obtained by \cite{jsk+08} are broadly similar to the model we produced in our analysis of \chimepsr{} timing data. However, they did not meaningfully detect the Shapiro delay from their observations. \cite{jsk+08} instead derived an upper limit on the inclination of the orbit at $i < 47 \deg$, based on their non-detection of the Shapiro delay. Using this constraint, they estimated the masses of the pulsar and companion to be $m_\textrm{p}$ $<1.17$ M$_\odot$ and $m_\textrm{c}$ $>1.55$ M$_\odot$. 

In our work, we were able to significantly detect the signature of Shapiro delay from \Jfifteen{}, with the grid-based inclination of the orbit estimated to be $i = 49.6^{+1.6}_{-1.8}$ deg. The derived inclination angle is marginally higher than the upper limit placed by~\cite{jsk+08}. The low inclination angle suggests that the Shapiro-delay signature from the \Jfifteen{} system is weak, and is only detectable due to a combination of the significant apsidal motion and high-cadence sampling of our data over much shorter time span of less than 4 years.

The measured masses of the \Jfifteen{} system also show that the pulsar is the more massive object with $m_\textrm{p} = 1.470^{+0.030}_{-0.034}$ M$_\odot$, whilst the companion neutron star has a smaller mass of $m_\textrm{p} = 1.248^{+0.035}_{-0.029}$ M$_\odot$. The rotational properties of the pulsar suggest that it is a partially recycled pulsar and would be the first born neutron star in the system. The first born neutron star being the heavier object in a double neutron star system is consistent with most others that are observed~\citep[See ][and references therein.]{tkf+17, of16} \cite{tkf+17} studied the formation of such systems and suggest that the mass discrepancy between the neutron star is unlikely due to accretion of mass by the first born neutron star, but rather due to the progenitor of the second neutron star being stripped of its mass more significantly due to the first neutron star prior to the second supernova.

We compared our best-fit Keplerian elements with those published by \cite{jsk+08}, in order to detect or constrain variations between the two data sets. No significant changes in $x$ or $P_{\rm b}$ beyond 2-$\sigma$ were found after accounting for clock-correction differences between the two timing solutions. The limits we derived from this analysis -- $|\dot{x}| < 3\times10^{-14}$ and $|\dot{P}_{\rm b}| < 6\times10^{-12}$ -- are consistent with those determined from the longer data set studied by \cite{jsk+08}.

The accuracy and precision in our measurement of $\dot{\omega}$ are nearly identical to the values published by \cite{jsk+08}. It is worth noting that this result arises from our analysis despite the \chimepsr{} data set being a factor of $\sim$3.5 shorter in timespan than that analyzed by \cite{jsk+08}. This circumstance demonstrates that the \chimepsr{} instrument can produce high-accuracy snapshots of pulsar orbits on much faster timescales than possible elsewhere, owing to the nearly-daily cadence of our observations, and thus allow for the quickened resolution of orbital variations over time. As originally noted by \cite{jsk+08}, the statistical uncertainty in the observed apsidal motion is comparable in magnitude to variations that arise evolving orientation of the binary system due to proper motion \citep[e.g.,][]{kop96}. We therefore consider our measured $\dot{\omega}$ to arise purely from general-relativistic orbital motion. A combination of the \chimepsr{} data set with that published by \cite{jsk+08}, which will lead to substantially improved measurement of $\dot{\omega}$ and likely other relativistic variations, will be the subject of future work.

\subsubsection{\Jtwenty{}}
\Jtwenty{} is a pulsar recently discovered by the FAST Galactic Plane Pulsar Snapshot survey~\citep{hww+21}.~\cite{hww+21} noted that \Jtwenty{} is detected with significant acceleration, suggesting that the pulsar has a binary companion. Our independent observations revealed that the pulsar is in a 0.7-day long orbit, with a detectable Shapiro-delay signature that revealed a companion with m$_\textrm{c}$ $=0.90^{+0.14}_{-0.12}$ M$_\odot$. Considering the low eccentricity of the orbit and the mass of the companion, it is likely to be a high mass carbon-oxygen white dwarf. 

The relatively long rotation period of \Jtwenty{} ($P > 10$ ms) compared to typical millisecond pulsars, combined with the relatively heavy mass of the companion ($m_{\rm c} > 0.5$ M$_\odot$), suggests that it is part of the class of binary system known as the intermediate-mass binary pulsars~\citep[IMBPs]{clm+01}. IMBPs are thought to be formed from accretion of a main sequence companion of 4-8 M$_\odot$, in which the pulsar itself is partially recycled. \Jtwenty{} has also a larger eccentricity compared to typical pulsar binary systems of similar orbital period, as well as a low scale height of $0.13 < z < 0.17$ kpc, for an estimated distance to the pulsar of 1.6-2.0 kpc as predicted by the Galactic electron density models~\citep{cl02, ymw17}. These properties are consistent with the system being an IMBP.

An optical source with $r = 23.55, g-r = 1.43$, SDSS J202320.88+285345.3, is found in the Sloan Digital Sky Survey~\citep[SDSS,][]{yaa+00, aaa+22}, around 5'' from the timing position of \Jtwenty{}. However, the photometric properties of the source, specifically the SDSS $u-g, g-r$ and $r-i$ colors, are inconsistent with it being a white dwarf, using the white dwarf cooling model described by ~\cite{bwb95, hb06, ks06, tbg11, bwd+11, bda18, bbbf20}\footnote{\url{https://www.astro.umontreal.ca/~bergeron/CoolingModels/}}, even considering the Galactic extinction along the line of sight, which gives $r = 24.17, g-r = 1.23$ ~\citep{sfd98, sf11}. This suggests that the optical source is not associated with \Jtwenty{}. No sources are detected with 5'' of \Jtwenty{} in the Two Micron All Sky Survey~\citep[2MASS]{scs+06} and the Gaia mission~\citep{gaia16, gaia22} The non detection of the WD companion is expected, as it should have a r-magnitude of 25 in SDSS, smaller than the detection threshold of the surveys above, based on the r-magnitude of the detection made on a similar WD in another pulsar-binary system, PSR J1658$+$3630~\citep{tbc+20}.

\section{Conclusions}

By leveraging the near-daily observations conducted by CHIME/Pulsar, we were able to, for the first time, model the Shapiro time delay of the binary pulsars PSRs J0218+4232, J1518+4904 and J2023+2853. \ef{In the case of J1518+4904, we assumed that the observed aspidal motion was described by general relativity so that its significance could be used to constrain the component masses of the system.} These measurements allowed us to constrain to the masses of the pulsars in the binary systems to be $1.48^{+0.24}_{-0.20}$ M$_\odot$, $1.470^{+0.030}_{-0.034}$ M$_\odot$ and \cm{$1.50^{+0.49}_{-0.38}$ M$_\odot$} respectively. The measured pulsar masses of all three system are well within the known mass distribution of neutron stars~\citep{of16}.

We were also able to measure the mass of the companion neutron star to \Jfifteen{} to be $m_{\rm c} = 1.248^{+0.035}_{-0.029}$ M$_\odot$, and for the first time, model the binary properties of \Jtwenty{}, revealing a relatively high mass companion of \cm{$m_{\rm c} = 0.93^{+0.17}_{-0.14}$ M$_\odot$}. The measured mass, together with a near-circular orbit, suggest that the companion is likely a high-mass carbon-oxygen white dwarf, and that the system is an IMBP. The low mass of the companion to \Jtwo{}, a helium white dwarf of $m_{\rm c} = 0.179^{+0.019}_{-0.016}\textrm{ M}_\odot$, is consistent with predictions from models of MSP formation through long-term mass transfer within low-mass X-ray binary systems \citep[e.g.,][]{ts99}.

We also obtained an estimate in the change in orbital period of \Jtwo{} between the epoch of the observations from~\cite{pdd+19} and our observations. We found that the primary contribution of the orbital period change is due to variations in the Doppler shift from Shklovskii acceleration, and that \cm{the change corresponds to a distance to the pulsar of $d$ = ($6.7 \pm 1.0$) kpc, consistent with the measurement of $d$ = 6.3 kpc from the VLBI observation conducted by~\cite{dyc+14}.} No significant changes were observed in the \chimepsr{} estimates of the orbital elements of \Jfifteen.

Our success in obtaining measurements of the masses of these pulsars through Shapiro-delay modelling, using the near-daily observations from CHIME/Pulsar, has prompted us to observe other binary pulsar systems on a near-daily basis. These high-cadence observations will help constrain the masses of the objects in these systems in order to provide an even larger sample of pulsar masses and relativistic measurments. We also continue to observe the three pulsar binary systems in this study, in order to further constrain their properties. 

\ef{Future work on these sources will also investigate the DM variations and noise properties of the high-cadence CHIME/Pulsar data sets, which could slightly modify the precision of our mass and geometric measurements. Moreover, subsequent studies will compute and analyze ``wideband" TOAs in order to maximize arrival-time precision and reduce TOA data volumes for efficient modeling of timing phenonema.}

\acknowledgements{
We acknowledge that CHIME is located on the traditional, ancestral, and unceded territory of the Syilx/Okanagan people. We thank Marten van Kerkwijk for valuable discussion.

We are grateful to the staff of the Dominion Radio Astrophysical Observatory, which is operated by the National Research Council of Canada.  CHIME is funded by a grant from the Canada Foundation for Innovation (CFI) 2012 Leading Edge Fund (Project 31170) and by contributions from the provinces of British Columbia, Qu\'{e}bec and Ontario. The CHIME/FRB Project, which enabled development in common with the CHIME/Pulsar instrument, is funded by a grant from the CFI 2015 Innovation Fund (Project 33213) and by contributions from the provinces of British Columbia and Qu\'{e}bec, and by the Dunlap Institute for Astronomy and Astrophysics at the University of Toronto. Additional support was provided by the Canadian Institute for Advanced Research (CIFAR), McGill University and the McGill Space Institute thanks to the Trottier Family Foundation, and the University of British Columbia. The CHIME/Pulsar instrument hardware was funded by NSERC RTI-1 grant EQPEQ 458893-2014.

This research was enabled in part by support provided by the BC Digital Research Infrastructure Group and the Digital Research Alliance of Canada (alliance-can.ca).
}

\software{
\dspsr{} \citep{vb11}, \psrdada{} (\url{http://psrdada.sourceforge.net}), \psrchive{} \citep{hvm04, vdo12}, \presto{} (\url{https://github.com/scottransom/presto}), \tempo{} (\url{https://tempo.sourceforge.net}), \tempotwo{} \citep{hem06}
}

\bibliographystyle{aasjournal}
\renewcommand{\bibname}{References}

\bibliography{references}

\end{document}